# Pionic contribution in the Drell-Yan dilepton production cross section in p-Cu collision in the framework of the shell and Fermi gas models

R. Rostami and F. Zolfagharpour

*Department of Physics, University of Mohaghegh Ardabili, Ardabil, Iran*

Email: rezarostami62@gmail.com

Email: zolfagharpour@uma.ac.ir

**Abstract**

*To investigate the high mass dilepton production cross section produced due to the Drell-Yan process in hadronic collisions such as nucleon- nucleus, the valence and sea quarks distribution functions inside nucleus is used. In this study, in the framework of the shell and Fermi gas models, by adding quarks distribution functions of pions inside nucleus besides the quarks distribution functions of bound nucleons, the changes in the dilepton production cross section were investigated. For this reason, pionic contribution in the structure function of $^{63}Cu$ nucleus and its EMC ratio was first studied using the aforementioned models. Then, in the framework of the Drell-Yan process using GRV's nucleons and pions quarks distribution functions, the high mass dilepton production cross section in p-cu collision was calculated and compared with the available experimental data. The extracted results, based on the two mentioned models, were greatly the same and by considering the pionic contribution, the theoretical results were consistent with the experimental data.*

**Introduction**

The process of dilepton production in hadronic collisions such as proton–nucleus have been studied by many research groups[1-4]. There are various channel for dilepton production. Dileptons with low mass can be produced due to decay of vector mesons such as $\varphi$, $\rho$ and $\omega$[1]. The vector mesons, such as $J/\psi$ and $\psi'$, can produce dileptons with intermediate mass, and the source of high mass dileptons is the Drell-Yan process[2]. In deep inelastic hadronic interaction of nucleons with nucleus, collision occurs between incident nucleon and bound nucleon or to be more precise between their quarks. Due to these interactions, the dileptons can be produced via aforementioned channel. To investigate the dilepton production cross section due to the Drell-Yan process, it is necessary to calculate the distribution of quarks inside nucleus and obtain the dilepton production cross section from non-coherent sum of scattering of individual projectile quarks from individual target quarks [5, 6]. In 1983, the results obtained from the experiments carried out by the European research group on the scattering of muons on iron and deuteron nuclei showed that the structure function of free and bound nucleons and consequently their quark distribution functions are different. The change in the quark distribution functions and consequently the structure functions of the bound nucleons to free is known as the EMC effect. The main contribution in the EMC effect and formation of the difference between the structure function of free and bound nucleons is related to two effects: 1) Fermi motion and 2) binding energy. However, this difference is not only due to these effects, but various particles in the nucleus, such as the resonant particles like Delta, and the exchange particles such as pions, can contribute to its formation[7-11]. Hence, it is necessary to distinguish between the bound and free nucleons quark distribution functions. To distribute nucleons inside the hadrons, different models have been presented and each of them has recorded its own success [12-14]. In this study, using the shell [15] and Fermi gas models[16] and considering the pionic contribution, as well as the Fermi motion and binding energy, the pionic contribution in the structure function of copper nucleus was first examined. Thereafter, the quark distribution functions of bound nucleons and pions were extracted in the shell and fermi gas models framework using the GRV's distribution functions of quarks in free nucleons [17] and distribution functions of quarks in free pion[18]. Finally, using these functions, the pionic contribution in the production cross section of the dilepton produced as a result of the Drell-Yan process was investigated.

**Calculation of the distribution function of bound nucleons using the Fermi gas and shell models**

To calculate the nucleons contribution in the structure function of nucleus, it is necessary to calculate their distribution function inside nucleus. This distribution function in the Fermi gas model can be expressed as follows [16]:

$$f_N^A(z)_F = A\left(\frac{3}{4v_F}\right)\left[1 - \frac{(z - \langle z \rangle_N)^2}{v_F^2}\right]\theta(v_F - |z - \langle z \rangle_N|) \tag{1}$$

in which $v_F = \frac{p_F}{m_N}$, $\langle z \rangle_N = 1 + V\left(m_N + \frac{3p_F^2}{10m_N}\right)^{-1}$. z represents a fraction of the total momentum of nucleus carried by nucleon and $v_F$ and $p_F$ are the speed and momentum of the Fermi gas, respectively and $m_N$ is the mass of nucleon. $<z>_N$ is the momentum carried by nucleons, while $V$ is the potential well depth of the nucleus. Fermi gas momentum is calculated as follows [19]:

$$p_F = 1.2525 - 0.1515 \log\frac{40}{A - 5.9}(fm^{-1}) \tag{2}$$

Where A is the atomic number of the nucleus. In the shell model, by considering the quark structure for bound nucleons inside the nucleus, their distribution function can be expressed as follows:

$$f_N^A(z)_{nl} = \frac{1}{2}\left(\frac{m_N}{\hbar\omega}\right)^{1/2}\frac{n!}{\Gamma(n+l+\frac{3}{2})}\sum_{t_1=0}^{n}\sum_{t_2=0}^{n}\frac{(-1)^{t_1+t_2}}{t_1+t_2}\begin{pmatrix}n+l+\frac{1}{2}\\n-t_1\end{pmatrix}$$

$$\times \begin{pmatrix}n+l+\frac{1}{2}\\n-t_2\end{pmatrix}\Gamma\left[l+t_1+t_2+1,\frac{m_N}{\hbar\omega}\left(z-1-\frac{\varepsilon_{nl}}{m_N}\right)^2\right] \tag{3}$$

Where $\varepsilon_{nl}$ is the mean one nucleon separation energy or on the other hand, is the average removal energy that has been considered the same for nucleons in different levels [15]. In this study, $\varepsilon_{nl}$ was considered to be different for nucleons in different levels.

$f_N^A(z)_{nl}$ should satisfy the normalization rule:
$$\sum_{N=n}\sum_{nl}\int_0^\infty dz g_{nl}^N f_N^A(z)_{nl} = A \tag{4}$$
$\hbar\omega$ in the neutral unit could be expressed as:
$$\hbar\omega(\text{MeV}) = \frac{42.2}{<r^2>_{nl}}\left(2n + l + \frac{3}{2}\right) \tag{5}$$
Where $<r^2>_{nl}$ is the mean square radius of the state *n, l* and its unit is Fermi.

**Calculation of the structure function of nucleus and its EMC**

The structure function of nucleus, by considering the pionic contribution in the shell and Fermi gas models, is expressed as follows, respectively [15, 20-22]:

$$F_2^A(x,Q^2) = \int_x^A f_\pi^A(z) F_2^\pi(\frac{x}{z},Q^2) \, dz + \sum_{N=n,p} \sum_{nl} \int_x^A dz \, g_{nl}^N \, f_N^A(z)_{nl} \, F_2^N(\frac{x}{z},Q^2) \quad \text{(for shell model)} \qquad (6)$$

$$F_2^A(x,Q^2) = \int_x^A f_\pi^A(z) F_2^\pi(\frac{x}{z},Q^2) \, dz + \int_x^A \frac{1}{2} f_N^A(z)_F \, (F_2^{N=n}(\frac{x}{z},Q^2) + F_2^{N=p}(\frac{x}{z},Q^2)) \quad \text{(for Fermi gas model)} \qquad (7)$$

In both equations (6) and (7), the first term is related to pionic contribution and the second term is related to nucleons contribution. $z = \frac{p_{nl}q}{m_N q_0}$, represents a fraction of the total momentum of nucleus carried by nucleons, where $p_{nl}$ is the nucleon momentum in the state $n, l$, $q$ is the momentum transferred from the lepton to the nucleon and $q_0$ is the energy transferred from the lepton to the nucleon, while $x = \frac{Q^2}{2m_N q_0}$ is the Bjorken scaling variable, where $Q^2$ is the square of the four-momentum transfer from the lepton. In the second term of equation (6), the first and second sums are related to the number of neutrons and protons and the quantum number of each energy state, respectively. $g_{nl}^N$ is the occupation number of the energy state $\varepsilon_{nl}$, in which for protons, N = p and for neutrons, N = n. $F_2^{N=n}(\frac{x}{z})$ and $F_2^{N=p}(\frac{x}{z})$ are the structure function of the free neutron and proton, respectively. In this study, the GRV's free neutron and proton structure function were used. $F_2^\pi(\frac{x}{z},Q^2)$ is the free pion structure function. The distribution function of pion inside the nucleus is given by the following equation [22,23]:

$$f_\pi^A(z) = \frac{3g^2}{16\pi^2} \Delta \lambda z \left[ \frac{1}{\lambda} \exp\left[-2\lambda \frac{t_0 + m_\pi^2}{m_\pi^2}\right] + \frac{1}{2} E_i \left[-2\lambda \frac{t_0 + m_\pi^2}{m_\pi^2}\right] \right] \qquad (8)$$

Where $E_i(-z) = -\int_z^\infty dt \frac{e^{-t}}{t}$, $t_0 = \left| m_N^2 \frac{z^2}{1-z} \right|$, $m_\pi = 139.570 MeV$ is the pion mass and $g = 13.5$ is the coupling constant. The cut off parameter λ plays the most substantial role, which shows changes in nuclear environments. Considering the pionic contribution and ignoring the other particles, the total momentum carried by the nucleons and pions can be written as follows:

$$<z>_N + \eta_\pi = 1 \qquad (9)$$

Where $<z>_N$ and $\eta_\pi$ are the momentum carried by the nucleons and pions, respectively and are defined as follows:

$$<z>_N = \frac{1}{A} \int_x^A dz \, z \, f^A(z) \qquad (10)$$

$$\eta_\pi = \int_0^{\frac{M_A}{m_N}} dz \, z \, f_\pi^A(z) = \frac{-V}{m_N} \qquad (11)$$

in which $M_A$, $m_N$ and v represent the mass of nucleus, the mass of nucleons and the potential well depth of nucleus, respectively.

The EMC ratio, which is defined as the ratio of the structure function of nucleus per nucleon to the deuterium structure function per nucleon, is given as:

$$R_{EMC}(x) = \frac{2F_2^A(x,Q^2)}{AF_2^{2H}(x,Q^2)} \qquad (12)$$

**Drell-Yan process**

Christenson et al. [3] studied the process of lepton pairs production in hadronic collisions for the first time and demonstrated its ability in order to provide a better understanding of the internal structure of hadrons. This reported process given in equation (13) is shown in Figure 1.

$$h_A + h_B \longrightarrow l^+l^- + x \tag{13}$$

This electromagnetic process is currently one of the most fascinating topics in high energy physics, which allows the comparison of theoretical ideas with very precise experimental data. The first model to produce a lepton pair was proposed in the seventies AD by Drell and Yan [4]. According to this model, a quark of a hadron is annihilated with an antiquark of other hadrons and a virtual photon is generated. The generated photon is converted into lepton pairs in hadronic collisions with opposite sign such as $\mu^+\mu^-, e^+e^-$. To calculate the differential cross section of lepton pairs production in collision between A and B hadrons, it is necessary to calculate the valence and sea quark distribution functions inside these hadrons. This production cross section in terms of quarks and anti-quarks distribution function is calculated as follows [24]:

$$\left(s\frac{d^2\sigma}{d\sqrt{\tau}dy}\right) = \frac{8\pi\alpha^2}{9\tau^{\frac{3}{2}}} k \sum_i e_i^2 \left[q_i^A(x_1)\bar{q}_i^B(x_2) + \bar{q}_i^A(x_1)q_i^B(x_2)\right] \tag{14}$$

Where $\sqrt{\tau} = \frac{m}{\sqrt{s}} = \sqrt{x_1 x_2}$, $\sqrt{s}$ is the hadron-hadron c.m. energy and $m = \sqrt{x_1 x_2 s}$ is the mass of dileptons. $e_i$ is the fractional electric charge of the quark $i$. $\bar{q}_i^B(x)$ is the ith anti-quark distribution function inside bound hadron B and $q_i^A(x)$ is ith quark distribution function inside the bound hadron A, which is defined as [25]:

$$q_i^A(x) = \sum_{nl}\int_x^A \frac{dz}{z} g_{nl}^N f_N^A(z)_{nl} q_i\left(\frac{x}{z}\right) + \int_x^A \frac{dz}{z} f_N^\pi(z)_{nl} q_i^\pi\left(\frac{x}{z}\right) \quad \text{(for shell model)} \tag{15}$$

$$q_i^A(x) = \int_x^A \frac{dz}{z} f_N^A(z)_F q_i\left(\frac{x}{z}\right) + \int_x^A \frac{dz}{z} f_N^\pi(z)_F q_i^\pi\left(\frac{x}{z}\right) \quad \text{(for Fermi gas model)} \tag{16}$$

Where $q_i(x)$ and $q_i^\pi(x)$ are the distribution functions of quarks inside the free hadrons[17] and free pions [18]. k is the correction coefficient known as k factor that is considered in order to provide an agreement between the experimental and the theoretical results.

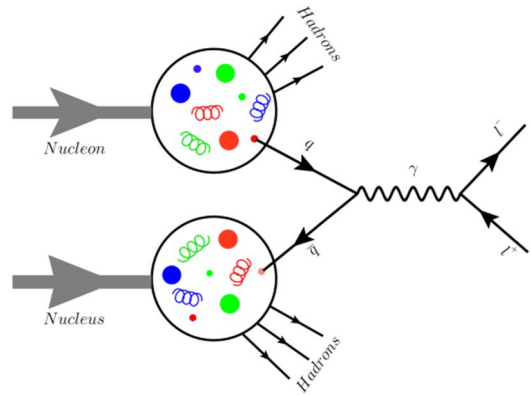

Figure 1. Drell-Yan process; a quark with fraction of momentum $x_1$ in the hadron A and an anti-quark with fraction of momentum $x_2$ in the hadron B collide and annihilate to a photon. This virtual photon with mass $m = \sqrt{x_1 x_2 s}$ subsequently decays into a lepton pair.

**Results and discussion:**

According to equations (6) and (7), to calculate the share of nucleons in the structure function of nucleus, the distribution function of nucleons inside the nucleus is required in addition to the free nucleon structure function. For this purpose, the GRV's free neutron and free proton structure functions were applied and using the parameters listed in Table 1 and according to equations (1) and (3), the nucleons distribution function in terms of the Fermi gas and shell models were calculated. In the Fermi gas model, the potential well depth is considered as v=-39 MeV. Figure 2 shows the GRV's free neutron and proton structure functions, which were applied in the calculations.

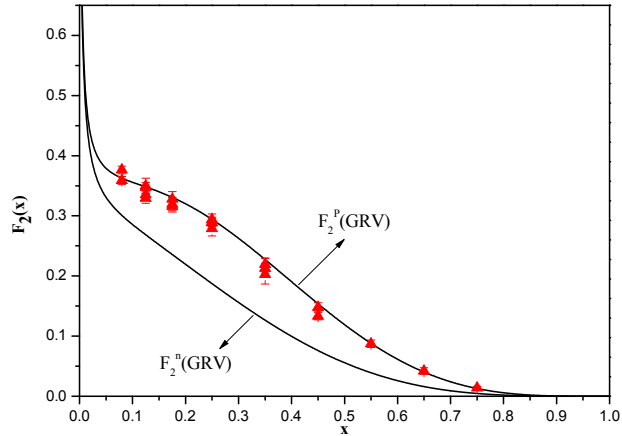

Figure 2. GRV's free neutron and proton structure function. The experimental data shown for free proton structure function are from [26].

Table 1: The variables used in calculating the structure function of $^{63}$Cu nucleus. The number in parentheses from left to right are ($< r^2 >^{1/2}$ (Fermi) , $\hbar\omega$(MeV), $g_{nl}^p$ , $g_{nl}^n$ , $\varepsilon_{nl}$(MeV)) parameters, respectively. $< r^2 >^{1/2}$ for each level is taken from [27] for each level.

| Nucleus | | |
|---|---|---|
| shell | $^2$H | $^{63}$Cu |
| 0s | (2.09,15.35,1,1,-1.5) | (1.67,22.23,2,2,-33) |
| 0p |  | (2.44,17.34,6,6,-32) |
| 0d |  | (3.10,12.51,10,10,-31) |
| 1s |  | (3.48,11.95,2,2,-30) |
| 0f |  | (3.95,11.92,9,14,-29) |

To apply pionic contribution in the structure function of nucleus in addition to pion structure function, the distribution function inside the nucleus should be calculated. For this purpose, the GRV's pion structure function was applied and according to equation (8), the pion distribution function inside the nucleus was calculated. In this study, $\lambda = 0.026$ and $m_\pi = 139.570$ MeV were used, which were taken from [22]. Using equations (9) and (10), $<z>_N$ and $\eta_\pi$ were calculated as 0.9675 and 0.0325 for Cu nucleus, respectively. convenient $\Delta\lambda$ for $^{63}$Cu and $^2$H nuclei was considered 0.00792 and 0.00242, respectively to be satisfactory for equation 11. Figure 3 shows the distribution function of pion inside

$^{63}$Cu and $^2$H nuclei. As shown in the figure, the distribution function of pion is increased for heavier nucleus. The distribution function of pions, which carry a very low or high fraction of the total momentum of nucleus, tends to zero. For both $^{63}$Cu and $^2$H nuclei in about $z = 0.35$, the distribution function of pions has a peak. In Figure 4, the GRV's LO and NLO pion structure function, as well as the structure function of pion calculated according to the equation $F_2^\pi(x)$ [29] are plotted. The GRV's LO pion structure function was used in the calculations.

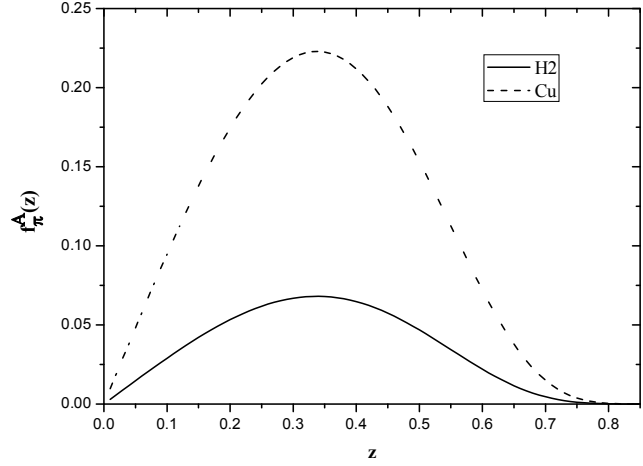

Figure 3. The distribution function of pion inside $^2$H and $^{63}$Cu nuclei.

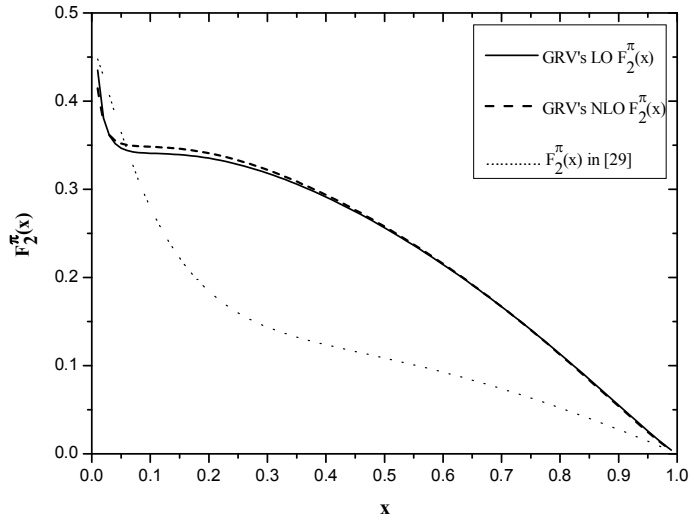

Figure 4. The structure function of pion. The full and dashed curve are the GRV's LO and NLO pion structure function, respectively. The dotted line is the structure function calculated according to equation $F_2^\pi(x)$ given from [29].

In Figure 5, the $^{63}$Cu structure function per nucleon in mean $Q^2 = 5 GeV^2$, in terms of the Fermi gas and shell models distribution functions of nucleons, is shown. Figure 6 shows the EMC ratio of $^{63}$Cu based on the $^{63}$Cu structure functions in both models. In both aforementioned figures, the full curve is obtained by considering the Fermi motion, the binding energy and the pionic contribution effect. The dashed line shows the structure function by considering the Fermi motion, and the binding energy. The upper and lower full and dashed curve were plotted based on Fermi gas and shell models, respectively.

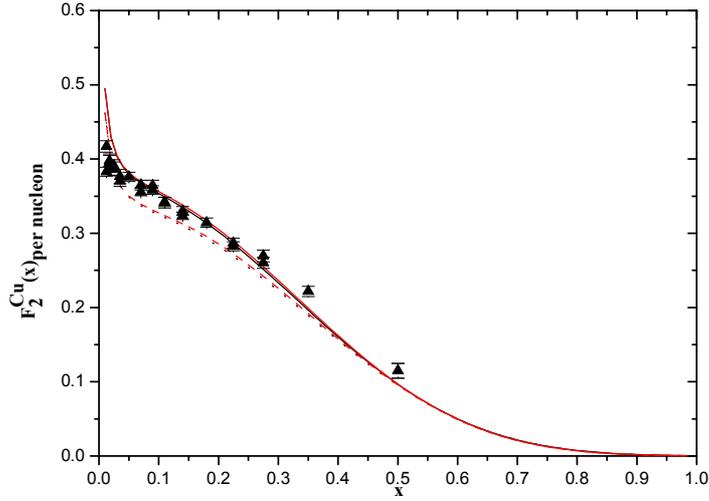

Figure 5. $^{63}$Cu structure functions per nucleon in mean $Q^2 = 5 GeV^2$. The full and dashed curve is obtained by considering, the Fermi motion, the binding energy, and the pionic contribution effect and without considering pionic contribution effect. The upper and lower full and the dashed curve were plotted based on Fermi gas and shell models, respectively. The experimental data shows deuterium structure function per nucleon [26].

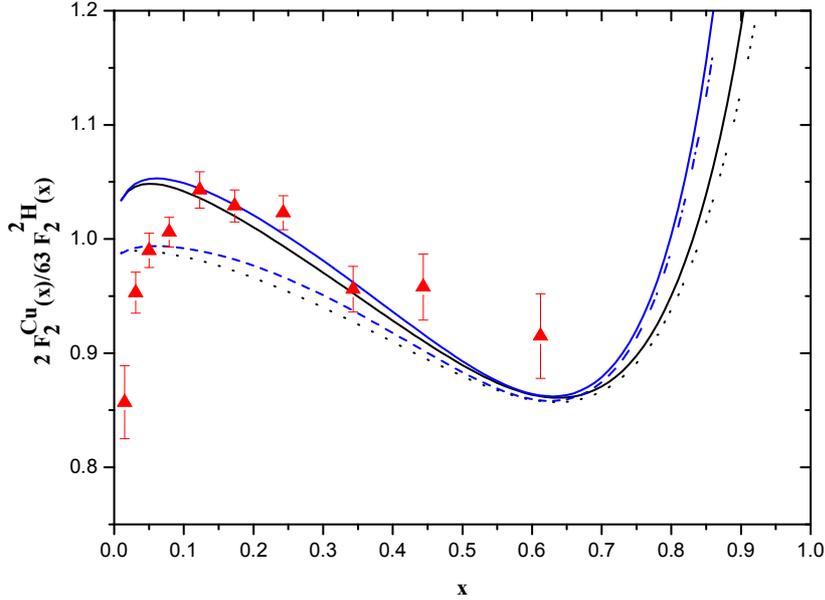

Figure 6. The ratio $R = \frac{2F_2^A(x,Q^2)}{AF_2^{2H}(x,Q^2)}$ in terms of x for Cu nucleus. The full and dashed curve were obtained by considering the Fermi motion, the binding energy, and the pionic contribution effect and without considering pionic contribution effect. The upper and lower full and dashed curve were plotted based on Fermi gas and shell models, respectively. The experimental data are from [26].

Figures 5 and 6 show that drawn lines, based on both models, are consistent. The results showed that in both models for x=0.15, the pion cloud increased the structure function of $^{63}$Cu to about 8% and its corresponding EMC ratio to about 5.5%. According to the result obtained by [20] and the extracted results from Figures 5 and 6, based on both models, most contribution in the $^{63}$Cu structure function and its EMC ratio is related to Fermi motion. The role of binding energy in small x is not considerable, but as x increases gradually, its contribution in mean x is enhanced and again as x increases, its contribution is decreased. However, the role of pionic contribution in small x is considerable in comparison with the role of binding energy, but in mean x, its contribution is about zero and is not comparable with the binding energy effect. Finally, in large x, pionic and binding energy contribution in the $^{63}$Cu structure function and its EMC ratio is the same and about zero. In Figure 7, using GRV's quark distribution functions, the distribution function of valence u quark is shown based on both models. The bottom full line shows the distribution function of valence u quark inside free nucleons. The middle and high solid lines show its distribution function inside $^{63}$Cu nucleus respectively in Fermi gas and shell model that are in agreement.

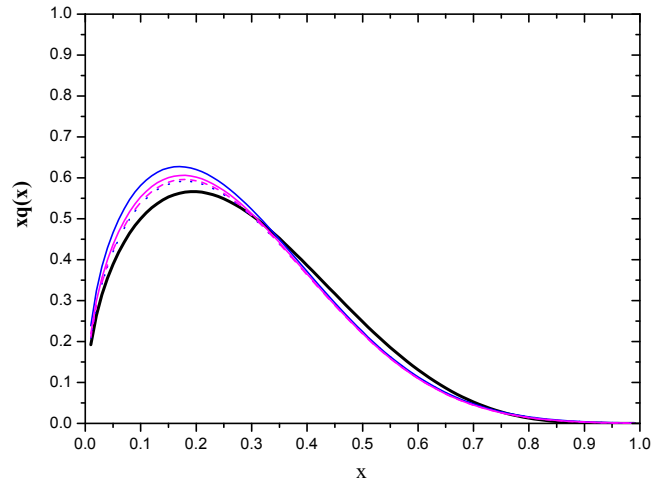

Figure 7. The distribution of valence u quark inside free nucleons and inside bound nucleons in $^{63}$Cu calculated using Fermi gas and shell models. The solid lines from bottom to up show the distribution of valence u quark inside free nucleons and inside $^{63}$Cu nucleus based on the Fermi gas and shell models. The solid lines were plotted by considering pions, and dotted lines were plotted without poins consideration.

Although, the dileptons are decayed from several channels, it is difficult to product the dileptons with low mass (*m* < 5 GeV/c$^2$) from the Drell-Yan channel. In this study, by applying $\sqrt{s}$ = 38.8 GeV and $\sqrt{\tau}$ from 0.1831 to 0.4359, the mass of dileptons was calculated from 7.10 to 16.91 GeV. Therefore, the source of the dileptons studied in this work is the Drell-Yan process. In Figure 8, by considering pions contribution, the dilepton production cross section produced as a result of the Drell-Yan process in p-$^{63}$Cu collision, using Fermi gas and shell models, has been plotted for y=0, 0.4. k factor has been shown in each graph and its amount increases with increasing y as shown in equation (17):

$k = \frac{3}{4}y + 1.7$ (17)

Different amounts of k factor have been reported by many research groups [24].
As shown in Figure 8, the plotted figures using the Fermi gas and shell models are in appropriate agreement and with the application of pionic contribution in both models, the dilepton production cross section was increased to about 3-8%.

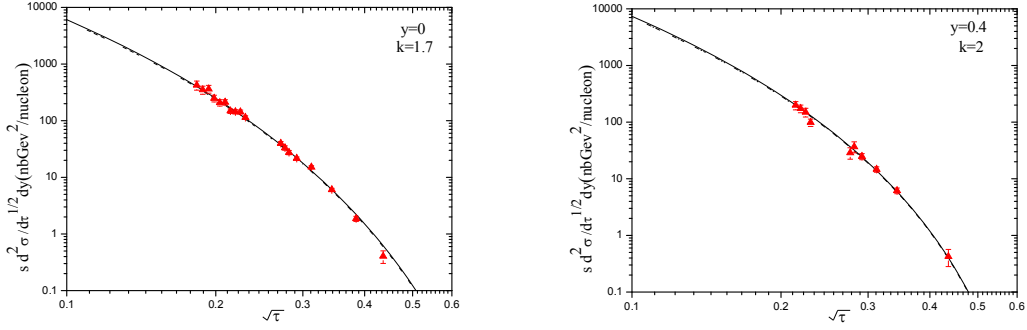

Figure 8. Dilepton production cross section for p-Cu collision in terms of $\sqrt{\tau}$ for y = 0, 0.4 and $\sqrt{s}$ = 38.8 GeV and in the application of pionic contribution. The solid and dashed lines were plotted using the shell and Fermi gas models, respectively. The used k factor was indicated in each graph. The experimental data are from [26, 28].

Figure 9 shows the dilepton production cross section for p-Cu collision in terms of $\sqrt{\tau}$ for y = -2 and $\sqrt{s}$ = 38.8 GeV. The dotted, dashed and full lines were plotted considering the Fermi motion alone, the Fermi motion together with the binding energy, and the Fermi motion together with the binding energy and the pionic contribution effect, respectively. As shown in Figure 9, in the production of dileptons, the most basic share is related to the effect of Fermi motion, which gradually decreases with increasing $\sqrt{\tau}$. Although, the pionic cloud do not have significant effect in the production of dileptons, but in small amounts of $\sqrt{\tau}$, its contribution (about 3-8%) in comparison with the binding energy (about 0%) is more prominent. Thus, with increasing $\sqrt{\tau}$, pionic contribution in the production of dileptons decreases gradually and tends to zero.

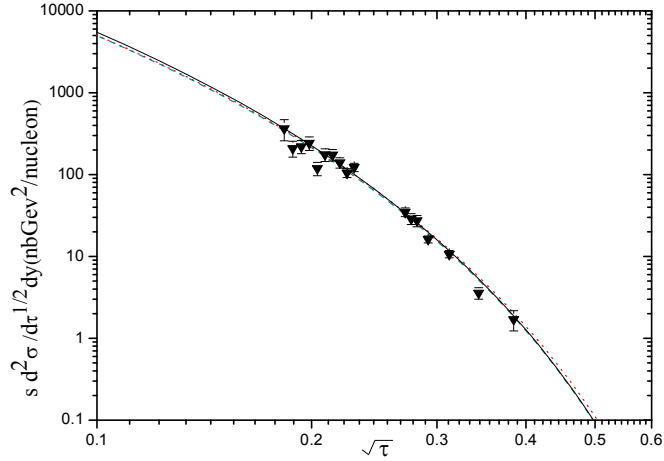

Figure 9. Dilepton production cross section for p-Cu collision in terms of $\sqrt{\tau}$ for y = -2 and $\sqrt{s}$ = 38.8 GeV. The dotted lines were plotted considering the Fermi motion. The dashed lines were plotted considering the Fermi motion and the binding energy. The full lines were obtained by considering the Fermi motion, the binding energy, and the pionic contribution effect. The experimental data are taken from [26,28].

**Conclusions**

In this study, using the GRV's free nucleons and pions structure functions, as well as the distribution functions of bound nucleons and pions inside $^{63}$Cu nucleus, the structure function of this nucleus was calculated. The distribution function of bound nucleons, based on the Fermi gas and shell models, was calculated and the structure function of $^{63}$Cu nucleus was obtained based on both models. The results showed that, based on both models, most contribution in the $^{63}$Cu structure function and its EMC ratio is related to the Fermi motion. In both models, for small x, by considering the pionic contribution, the structure function of this nucleus and its EMC ratio increased 8 and 5.5%, respectively and the pionic contribution was considerable in comparison with the binding energy, while in mean x, its contribution was about zero and as such, cannot be compared with the binding energy effect. Moreover, the GRV's distribution functions of quarks in free nucleons and distribution functions of quarks in pion were used to extract the quark distribution functions of bound nucleons and pions inside $^{63}$Cu nucleus based on both models. Finally, these functions were used to calculate the pionic contribution in the Drell-Yan dilepton production cross section for p-Cu collision in terms of $\sqrt{\tau}$ in various y and in $\sqrt{s} = 38.8$ GeV. The results showed that, based on both models in production of dileptons, the most basic share is related to the effect of Fermi motion. However, despite the low share of the pionic cloud in the production of dileptons, when applied, the theoretical results showed more consistent with the experimental results. In small amounts of $\sqrt{\tau}$, the dilepton production cross section increased to about 3 to 8%, which was more prominent in comparison with negligible share of binding energy.